\documentclass{article}

\usepackage{amsmath}
\usepackage{amssymb}
\usepackage[mathcal,mathscr]{eucal}
\usepackage{eufrak}
\usepackage{epsfig}
\usepackage{amsthm}
\usepackage{graphicx}
\usepackage[numbers, sort&compress]{natbib}

\begin{document}


\begin{center}
\Large{\textbf{Quantum Mechanics, is it magic?}\footnote{On April, 13 professor John A. Wheeler passed away. We express here our admiration for the work he has legated and we dedicate this paper to honour his memory.}}
\end{center}

\begin{center}
M.\ Ferrero\\
Dpto.\ F\'{\i}sica, Universidad de Oviedo, Spain\\
\texttt{maferrero@uniovi.es}

\medskip

D. Salgado\\
Dpto. Ingenier\'{\i}a Inform\'{a}tica, Universidad Antonio de Nebrija, Spain\\
\texttt{dsalgado@nebrija.es}

\medskip

J.L. S\'{a}nchez-G\'{o}mez\\
Dpto.  F\'{\i}sica Te\'{o}rica, Universidad Aut\'{o}noma de Madrid, Spain \\
\texttt{jl.sanchezgomez@uam.es} \\
\end{center}

\medskip


\vspace{1cm}

\begin{center}
\begin{minipage}{10cm}
We show that quantum mechanics is the first theory in human history that violates the basic \emph{a priori} principles that have shaped human thought since immemorial times. Therefore although it is more contrary to magic than any body of knowledge could be, what could be called its magic precisely resides in this violation.\end{minipage}
\end{center}

\vspace{1cm}


\section{Introduction: the use of the word magic in quantum physics}
This paper has been inspired by all those physicists that at a certain point have used the word magic when describing, explaining or writing about Quantum Physics. What called our attention was the somewhat anachronic use of the term. From a rigorous anthropological point of view, which is the context were the word magic can properly get its meaning, magic is a previous stage to religion and science in the evolution of human thought. Although it could be said that science and magic share some common aspects and could even coexist within different groups in a concrete period of time, they are essentially different institutions created by the human beings. To point out this difference is one of the aims of this paper. The other will be to show where the magic of quantum mechanics resides.\\

The word magic has three different meanings. According to the first one, magic is "the pretended art of influencing the course of events by compelling the agency of spiritual beings or by bringing into operation some occult controlling principle of nature". The second is "a secret and over-mastering influence, resembling magic in its effects". The third is "the art of producing (by legerdemain, optical illusion, etc.) surprising phenomena resembling the results of magic" \cite{EngDicHisPri88a}. As the second and third definitions include the very same word magic, the use in these senses necessarily requires some kind of previous knowledge of the meaning of magic in the first sense.\\

It is true that since its inception, Quantum Theory, perhaps the most effective physical theoretical structure built-up by human beings, has had a shaky philosophical foundations, giving way to many discussions between the founder fathers of the theory (N. Bohr, A. Einstein, W. Heisenberg, E. Schrödinger, M. Born, W. Pauli and P. Dirac). The use of words like paradox, puzzle, mystery, extraordinary, weirdness, spooky, etc, to manifest the surprise that some of the theory's predictions caused upon us, to wit, mainly those defying our intuition and common sense, were frequent. However, the word magic that, as we shall see very soon has more profound implications entered the scene only at a later stage in the development of the discussion.\\

As far as we know, the first time the word magic was written down in the context of "hard physics" was 1972, in a book entitled \emph{Magic without Magic}, edited by J. E. Klauder to honour Professor John Archibald Wheeler in his sixtieth anniversary \cite{Kla72a}. The sense in which the word was employed in the title of this book is not specifically explained within it. Nevertheless the title can speak for itself. If we consider which one of the three aforementioned meanings of the word magic can be applied to the person to whom it was dedicated, the life and work of J. A. Wheeler could be taken as the perfect example of the second meaning of the word. This is not a capricious statement. It is a constant in the opinions of all those that have been lucky enough to meet or have any type of relationship with him. Let us quote a representative sentence of the aforementioned book to briefly remember his character and the role he has played in XX century physics. "The innovative feature of John Wheeler is remarkable throughout his career as a theoretical physicist. In field after field, he has generated the key concepts that have been used by others. He was the pioneer who has opened up valuable new terrain and pointed the way. Many of those who later work in these fields do not know that it was John Wheeler who started them, because his great modesty, informality, and continual willingness to give credit to others have often efface his own germinal role" \cite{Tol72a}. In short, John Wheeler has had an "over-mastering influence, resembling the magic in its effects": he was, presumably, magical without magic (briefly and in French: J. Wheeler is \emph{magique sans magie}).\\

We had to wait another decade to find this word used in physics, but this time the sentence in which it was included became famous: "quantum mechanics is magic". Daniel Greenberger introduced this celebrated sentence in the discussion remarks at the Symposium on \emph{Fundamental Questions in Quantum Mechanics} held at the State University of New York, Albany, in April 1984 \cite{Gre}. The sentence immediately called the attention of the scholars working in the Foundations of Quantum Physics and it definitely achieved fame one year later when David Mermin open with it a nowadays classic paper on Foundations of Quantum Mechanics published in \emph{Physics Today} in 1985 \cite{Mer85a}.  In the last page Mermin affirms: "The EPR experiment is as close to magic as any physical phenomenon I know of, and magic should be enjoyed". The same D. Greenberger repeated the formula throughout the years. In the preface of the book \emph{Fundamental problems in Quantum Theory}, published to honour Professor J. A. Wheeler in 1995, he writes: "What makes quantum mechanics so much fun is that its results run so counter to one's classical intuitions, yet they are always predictable, even if unanticipated. That is why I like to say that quantum mechanics is magic, but it is not \emph{black} magic" (his emphasis) \cite{Gre95a}. In the collection of statements gathered on the occasion of the meeting \emph{Quantum Physics of Nature}, held in Vienna ten years later, Greenberger says again: "Quantum mechanics is magic! It is not black magic, but it is nonetheless magic!" \cite{Ard05}.\\

Putting aside the distinction introduced by Greenberger with the expression "black magic", and using the word magic in its third meaning, we must agree with him. The results of experiments like the double slit experiment carried out at low intensity \cite{VarExp}, the delayed choice experiment \cite{Whe83a} and many other experiments carried out in the last 35 years in Foundations and Quantum Information Processing (QIP) \cite{VarExp2} show that the behaviour of individual quantum systems is not only fun or enjoyable: it is indeed amazing!\\

If that were all, it would have been useless to pay any attention to the use of the word magic in the context of quantum physics. Almost everybody is using it in the correct \emph{third meaning}. However, this consideration does not close all the questions. Modern physics was born against common sense and has always been amazing for both the lay man on the street and for those who had enough knowledge of it. What are the reasons behind stating that quantum mechanics is magic, while the same expression is not applied, for example, to relativity, classical mechanics, geology, chemistry or biology? All of them are fun, enjoyable, and lead to results that are unexpected and surprising for the laymen. Why is quantum mechanics different in this respect? Where does its strangeness reside? Why its results seem to be paradoxical or extraordinary?\\

 We thought that if one considers the word magic from the point of view of its first basic meaning (not contemplated until now) we can obtain a significance which exceeds by large the particular realm of physics. It has much more physical and philosophical interest and, at the same time, it could give us a clue about these previous queries. In this paper we address the question of what the full implications of the statement "quantum mechanics is magic" are, when taking the word magic in its first foundational sense. We shall deal with these issues without pretending to present a final and "real truth". But rather to express a viewpoint that we believe could be satisfactory in order to dissipate the confusion around the term magic and to elucidate upon what this term can really contribute with in order to understand that the profound reason behind its use is that quantum mechanics is a theory that possesses a completely different structure, that is a different body of scientific knowledge to all previous ones. This is meant in the sense that it does not require the introduction of \emph{a priori} principles regarding nature that other bodies of knowledge implicitly introduce. The violation of those \emph{a priori} principles and the surprise that this violation causes upon us appears as the real reason behind the use of the word magic in quantum mechanics. This is an interesting result worth being considered.\\

The paper is organized with the following structure. In part two we make a rough approximation to the concept of magic in its anthropological sense. Part three briefly addresses the question of the similarities between magic and classical physics, with the foundations of all different sciences, modulo quantum physics. Part fourth intends to show how quantum entanglement implies a big break from the principles and foundations of both magic and other scientific theories. Part fifth contains the conclusions and some relevant considerations.

\section{An approximation to the concept of magic}
For the purpose of this paper we will adopt the characterization of magic introduced by Sir James Frazer, one of the founders fathers of modern anthropology, in his seminal work \emph{The Golden Bough} \cite{Fra93a}, a truly masterpiece of intellectual work. In page 11, under the title \emph{The Principles of Magic} Frazer says: "If we analyse the principles of thought on which magic is based they will probably be found to resolve themselves into two: first, that like produces like, or that an effect resembles its cause (Law of Similarity); and second, that things which have once been in contact with each other continue to act on each other at a distance after the physical contact has been severed (Law of Contact or Contagion)". Through the first principle a magician infers that she/he can produce any desired effect by imitating it; through the second, that whatever she/he does to a material object will affect the person with whom the object was once in contact. The first law produces 'Imitative Magic' and the second, produces 'Contagious Magic'. "If my analysis of the magician's logic is correct, its two great principles turn out to be merely two different misapplications of the association of ideas. Imitative magic is founded on the association of ideas by similarity; contagious magic is founded on the association of ideas by contiguity [\dots]. In practice the two branches are often combined [\dots] and may conveniently be comprehended under the general name of Sympathetic Magic, since both assume that things act on each other at a distance through a secret sympathy, \emph{\textbf{the impulse being transmitted from one to the other by means of what we may conceive as a kind of invisible ether}}, \emph{not unlike that which is postulated by modern science for a precisely similar purpose, namely, to explain how things can physically affect each other through a space which appears to be empty"} (p. 12, emphasis ours). Thus, although things might be separated, they are nevertheless always in contact through sympathy in such a way that whatever is done to one, must affect the other.\\

Let us introduce a simple "gedanken example" to explain these two laws further. Imagine a dog that always sees a bicycle lying aside near to the place where his owner frequently gives him a delicious piece of meat with some bone fragments. Imagine also that one day, after the sudden dead of his owner, lost, hungry and wandering around for food, he sees a similar bike laid by. What the magic laws say is that the dog will associate this bike with the other one and by contiguity with the delicious meal that he is now whishing and missing. (The hypothetical example could even be subject to empirical test by examining the mouth of the dog just after the encounter with the bike. If the hypothesis were correct, the mouth of the dog should be well salivated just after that encounter).\\

Frazer observes that the magician "tacitly assumes that the laws of similarity and contact are of universal application and not limited to human actions (p.11) [\dots] The magician does not doubt that the same causes will always produce the same effects, that the performance of the proper ceremony will inevitably be attended by the desired result [\dots] yet his power [\dots] is by no means arbitrary [\dots]. He can wield it only so long as he strictly conforms to the rules of his art or to what may be called the laws of nature as conceived by him [\dots] the succession of events is assumed to be perfectly regular and certain, being determined by immutable laws, the operation of which can be foreseen and calculated precisely; the elements of caprice, chance and accident are banished from the course of nature" (p. 49). The essential necessary characterization of magic for our purpose is already contained in which we have just said. This is all we need.\\

If the previous analysis done by Frazer is correct, underlying the whole system of magic is a kind of belief in the order and uniformity of nature. Magic assumes that in nature one event follows another necessarily (causality principle or \emph{principle of sufficient reason} plus the continuous flow of time: a causal explanation must respect the direction of time). And although Frazer says nothing about it, it is clear in his characterization that some other methodological principles also permeate the whole system of magic. One of them is the \textit{principle of realism}, that is, that the world is composed by objects that have properties that allow us to differentiate one from another. Just the implicit consideration of this principle gives sense to the two fundamental and specific laws of magic enunciated by Frazer. But not only that. Implicit in the second law it can also be perceived a \textit{separability} and \textit{locality principles} working in the magician's approach: once they have been in contact they continue to act upon each other by sympathy at a distance after the physical contact has been severed. In order to have such a situation, we need that things exist separately and individually. To argue it in the negative sense: the second law says that if two things had never been in contact, whatever is done to one would not affect the other whatsoever. That is, they are not only separate entities (separables, to use a modern term, see below), but also independent. Individuation is not lost through contact, or interaction, because once the contact has been severed, sympathy, "a kind of invisible ether", connects one with the other. This clearly implies a locality principle: things cannot be manipulated at a distance without the existence of any mediator.  Spooky actions at a distance, to quote Einstein, are not possible within magic. Note that the second law of magic says nothing about further and subsequent interactions. Implicit again in the system is the idea that a new contact does neither severe the previous one, nor establish new relations between the second and third that have never been in direct contact. Swapping would be impossible within magic.\\

To sum it up. As characterized by Frazer, the system of magic has five underlying principles and two laws. The principles are: the principle of realism, the causality principle, the separability principle, the locality principle, and the continuous flow of time in one direction. The two laws are: the law of similarity and the law of contact.

\section{Magic and the Foundations of Classical Physics}
As depicted in the previous paragraphs, magic appears to be a legitimate body of knowledge with close resemblances with the sciences in general and with physics in particular. We refer now to physics in a broad sense and in the period elapsed between Copernicus and the beginning of the Twenty Century. The foundations of these similarities are what we are going to elaborate now.\\

The first and most important resemblance is that both magic and classical physics believe in an established order in nature, determined by eternal and immutable laws that permit to foresee the course of events with certainty and which allow us to act in accordance with these. The succession of events is assumed to be perfectly regular and certain, determined by those eternal laws. The elements of caprice and chance are banished from the course of nature, as Laplace - in a well know sentence that we refrain from quote- superbly summarized. Therefore, the fundamental conception of magic is in this respect coincident with that of classical physics. The basic forces that govern the world are both impersonal and unconscious, as opposed, for example, to religion, that conceives of them as personal and conscious. "It is true that magic often deals with spirits, which are personal agents of the kind assumed by religion [\dots] but it treats them exactly in the same fashion as it treats inanimate agents [\dots]. Thus it assumes that all personal beings, whether humans or divine are in the last resort subject to those impersonal forces which control all things" \cite{Frabis}.\\

The second important similarity between magic and classical physics is that both satisfy the principles of realism, causality (sufficient reason), continuous time flow, separability and locality. It is true that the gravitation law in classical physics was non-local, and that through its influence, non-local actions reigned in physics for almost two hundred years. Yet it is also true that since it was introduced, even Newton perceived it as to be simply unintelligible, as a provisional expedient to be eliminated from physics. As it is well known, in a letter to Bentley dated 1693, he wrote how "It is inconceivable that inanimate brute matter should, without the mediation of something else which is not material, operate upon and affect other matter without mutual contact\dots That gravity should be innate, inherent, and essential to matter, so that one body may act upon another at a distance through a vacuum, without the mediation of anything else, by and through which their action and force may be conveyed from one to another, is to me so great an absurdity that I believe no man who has in philosophical matters a competent faculty of thinking can ever fall into it" \cite{Jan04a}. As we have seen, one way of avoiding the action at a distance and restore locality is to go back to magic, leaving open the possibility that gravity is transmited thanks to the presence of a kind of ether that penetrates and acts on all matter. It is well documented that Newton speculated about such ether as a medium for the gravitational interactions of material objects at least since 1679, as it can be read in a letter written to Boyle \cite{Janbis}. Seen in this context, the ravings of Newton with magic (alchemy) \cite{CommentNewton}, his search for the philosophers' stone to gain an understanding and power over nature in a historical moment in which the old medieval knowledge was becoming useless is not only reasonable, but it also takes a new and more plausible perspective.\\

This second similarity can be easily perceived by looking at the following frame:

\begin{table}[h!]
\begin{tabular}{|c|c|c|c|c|c|}\hline
    & Realism & Causality &  Separability & Locality & Flow of  time\\\hline
Magic  &    yes & yes & yes & yes & yes\\\hline
Classical Physics & yes & yes & yes & yes & yes\\\hline
\end{tabular}
\caption{\label{fram1}Frame 1. - Methodological principles working in magic and in classical physics. The yes/no inside the boxes are the answers to the question: Does this body of knowledge (magic/ c. physics) satisfy the principle of \dots?}
\end{table}

As we see in frame \ref{fram1} the same methodological principles underlie the two bodies of knowledge, magic and classical physics. However, classical physics is considered to be a genuine scientific theory, while magic is not. This comment sets out a very interesting question, namely: can we establish \emph{a priori} principles as necessary preconditions that then any physical theory must satisfy in order to be considered scientific? If the answer to this question is negative, as our previous comparison seems to suggest, then any argument based on \emph{a priori} principles introduced to criticize as unsatisfactory, incomplete or incorrect a scientific body of knowledge, is doomed to failure. The Einstein, Podolsky and Rosen (EPR in the following) argument \cite{EinPodRos35a} would be a paradigmatic example in this respect.\\

The essential point about what makes scientific a concrete body of knowledge must then reside somewhere else. The true failure of magic to be a science does not lie either in its general assumption about the order and uniformity of nature, in the implicit use of some \emph{a priori} principles or in the hypothesis that the events are determined by law. All these characteristic aspects are shared with classical physics. The fatal flaw lies in the wrong nature of the particular laws that govern the events. That is, whereas the rules laid down by the different sciences are derived by an inextricable combination of hypothesis and experiment, from ideas and careful experimentation or observation of phenomena, from conjectures and refutations to be more precise, "the order in magic is an extension by false analogy of the order on which ideas present themselves to our minds" \cite{Ref11}.  "The same principles which the magician applies in his practices are implicitly believed by him to regulate the operations of inanimate nature [\dots]. In short, magic is a spurious system of natural law [\dots] a false science" \cite{Ref11bis}.\\

The main difference between a scientific body of knowledge and a non scientific one such as magic, is therefore not located in the belief of an established natural order or in the introduction of \emph{a priori} principles, but in the specific laws and theories which govern events and thus in the method through which we obtain this knowledge. The magician \textbf{translates directly to nature} the laws that are present in his head, while in science the laws seem to be derived from \textbf{the previous order} that we humans have managed to introduce in a concrete field of the material reality (and not the other way round, that the order is a consequence of everlasting laws, see below). For the magician, the origin and legitimacy of the laws reside in our heads, while for scientific theories the origin and legitimacy reside in nature itself (through our intervention).\\

Let us emphasize with a well known example the point that we are trying to make. Aristotelian physics gave, for more than two thousand years, a perfect explanation of movements. This was due to his theory of natural places and violent movement. The specific laws of Aristotelian physics could be also classified themselves into two: first, that each object has a natural place in the universe to which it naturally returns if free to do so; and second, that any object not going to its natural place that has certain velocity must be moved by the force exercised by some other. Natural movement did not require any explanation, as this lies in the very nature of things themselves (for example, its different densities). All the other movements, characterized by the fact that the objects had some speed, were violent, and speed requires the presence of a force necessarily performed by another body: nature abhors vacuum. Even today not educated persons in terms of Newtonian physics usually give an explanation of movements in those Aristotelian terms. Yet, Aristotle's physics is not considered to be a scientific body of knowledge as we understand it today. Modern physics was born in the dawn of the XVII century when Galileo was able to refute this particular Aristotelian theory about the movement, introducing new specific laws masterly developed years later by Newton. Galileo changed Aristotle's natural movement to the inertial one (what is the explanation of the uniform movement?) and Newton showed that the force caused a change in the velocity, as the two first Newton's laws state. However, as in the case of magic, Aristotelian physics satisfied both the belief in an order and uniformity of nature and the \emph{a priori} principles mentioned above in frame 1, locality included. The flaw was, once again, in some other place, to wit, in his two specific laws about the movement.\\

To summarize, although we cannot enter now into detail, since immemorial times the evolution of human thought seems to have past first from survival to magic, then from magic to religion and finally from religion to sciences. The transition from magic to religion, not touched upon until now in this paper, occurred when it became clear (at least for the magician's elite) that the laws of magic didn't work properly. They became progressively conscious that they were not getting the results they were asking for. Then, in a clever move designed to maintain their privileges, the impersonal beings, subjects to the same laws than us, were substituted by some others that were able to violate these "immutable" laws. The following transition to science was significantly more painful and it took place when we went back to a different kind of immutable laws, when we dispensed with these (unsatisfactory) supreme gods and became "alone in the universe". "But magic, by conceiving a subjacent order subject to law directly prepared the way for science" \cite{Ref11tris}.\\

However, as it can be intuited from what we have said already, this historical development has taken place handed down from the remote antiquity of some fundamental principles that until the beginning of XX century had not been questioned, never. The transitions from one period to the other were realized without violating any of them. This allows us to understand why they are so deeply rooted in our minds.

\section{Quantum Entanglement}

We will try to show now how quantum mechanics produces a dramatic break with these principles for the first time in the history of human thought. To do it without using explicitly the quantum mechanical formalism, we will introduce a characteristic property of quantum systems known by the name of entanglement.\\

Entanglement is a formal concept deriving from two fundamental aspects of the quantum mechanical formalism. Namely, the superposition principle and the tensor product composition of the joint space state of two o more subsystems. It was, in some primary sense, introduced by EPR in 1935 in the very famous paper quoted above and published in the Physical Review \cite{EinPodRos35a}. Since its inception until nowadays, entanglement has been considered to be "the essence" of quantum mechanics \cite{Sch35a}. After a long process of development characterized by slow downs and accelerations, it set off in the nineties as the basic resource for information processing, and today it has become the central concept in quantum information theory (QIT), a new and emergent field in quantum physics with an outstanding potential to change our societies of information in the XXII century. In what follows, we will adopt the position that entanglement is a wholly physical property: it simply happens. It exists as a new aspect of nature, of material reality. This is not a voluntary or an arbitrary choice: it is based on the result obtained in many different experiments \cite{VarExp2} carried out in the last 36 years that converge in this very same aspect. Once more, material reality manifests itself in a far more complex way than it was previously thought, as it has always been the case in the history of physics. This renewable complexity is what makes, from our point of view, ontological completeness \cite{Mau07a} an oxymoron.\\

Entanglement is, on the other hand, the explanation for quantum correlations. Let us explain this a little further. There are correlations everywhere in the world: in classical physics, in our societies, in our personal relations, anywhere. This is the very reason why we can introduce order in nature. So, what is the problem with quantum correlations? Are they different from the others? The answer to this question is: yes, they are. Quantum correlations are stronger than classical ones and cannot be imitated by any classical correlation (Bell's theorem). At the philosophical level, we could say that classical correlations can be explained on the basis of the principle of sufficient reason, reduced to the conceptual level of cause and effect. That is, they can be understood on the basis of common causes (in the past) and/or hidden communication. Let us put an example. Imagine that two friends, Alice in London and Bob in Rome, are asked to produce one hundred times the "face side" (heads) or the "cross side" (tails) of an euro coin. This is a very simple experiment. What are the expected results? The results will be that Alice will get heads in -approximately- half of the cases (the same for Bob) and that in -again approximately- 25\% of the cases both of them will get heads and in other 25\%, tails. What would you think if they get \textbf{always}, in this and in many other experiments carried out equally, the same result (heads-heads or tails-tails)? Nobody would believe that this could happen by an implausible coincidence, by mere chance. Contrarily, and as an explanation of the persistent coincidences, we would seriously think that they have some device that allows both of them to get heads or tails at will as well as any of these two possibilities:\\

\begin{enumerate}
\item[1st. -] They had previously agreed in the order in which they were going to produce the face side or the cross side (common cause in the past), or
\item[2nd. -] They have a special equipment (that we do not see) that tells one of them the result that is being produced by the other (hidden communication).
\end{enumerate}

Any other alternative would be logically rejected, implying that these correlations could be reduced to the categories of cause and effect.\\
What about the quantum correlations? To make a long story short, in the particular case of experiments devised to verify the quantum correlations of maximally entangled pairs, we would \textbf{always} get heads-heads or tails-tails. However,
\begin{enumerate}
\item[1st. -] It may sound surprising or even absurd for those that do not know quantum mechanics, but the real fact is that "quantum coins" have neither heads, nor tails \textbf{until} they are observed, that is, registered or detected. This is again not an arbitrary decision. It has been empirically tested in many, many different and independent experiments carried out in the last one hundred years.
\item[2nd. -] No hidden communication exist (this possibility has been also excluded empirically and nowadays can be consider to be a well established experimental fact).
\end{enumerate}

When considered jointly, these two characteristics imply that to get such correlations what Alice does and obtains in London instantaneously has a consequence in Rome (it defines the corresponding value, face side or cross side, undefined until this very precise instant). On the other hand, and as Bell's theorem proves, no common cause is possible: there are not local hidden variable theories able to reproduce the quantum correlations.\\

Material reality is here manifesting quantum correlations without referents, without correlata \cite{Mer98a} (remember, the heads/tails do not exist until detected) in a holistic view (the result we get here has irretrievable "instantaneous" consequences there).\\

To avoid such amazing conclusion, it has been occasionally mentioned that what Alice does in London has not instantaneous consequences in Rome, because it was impossible for Bob to find out what Alice has been doing. He will keep getting 50\% heads and 50\% tails, and from these results it is completely unfeasible for him to ascertain if she has measured or not, and therefore the outcomes \cite{BraMet07a}. Strong correlations are useless when it comes to send instantaneous messages! This is absolutely correct. Quantum formalism guarantees that these correlations cannot be used to send messages at superluminal velocity. Today the agreement regarding this matter has made this discussion to come to an end. However, let us be very clear in this precise point. The problem is that we are not trying to elucidate upon that question. What we would like to emphasize is that, if Alice and Bob arrange afterwards to meet half way -let us say Paris- with their respective result notes (heads, heads, tails, and so on), they will confirm that they had \textbf{always} obtained the same result. This is what quantum mechanics predicts, verified by the experiments on Bell's inequalities, teleportation and cryptography. Moreover, it is precisely this absolute coincidence what we are interested in now. To be able or not to be able to use correlations to send signals faster than the speed of light can be a very important issue, but in our discussion is secondary \cite{Fra07a}. Let us emphasize it again: what we want to know is how it is possible that Alice and Bob obtain the same result when right before the ending of each test the "quantum coins" did not even have a definite state!\\

The answer to the different questions sorted out in the two previous pages about quantum entanglement is that the methodological principles of realism, causality, separability and locality are not satisfied in quantum theory. Let us explain further how this crucial break is produced. Realism, as it is usually understood in physics, is violated by entanglement because the "quantum coins" do not have a fixed value, but a range of probabilities. This is not a peculiarity of entanglement. In quantum theory there are not dispersion free states for all observables: non-commuting observables cannot have definite values simultaneously, as the uncertainty relations reflect. As a consequence, quantum systems, like our imaginary coins, have not in general properties with definite values \cite{CommentKS}. Causality is violated by entanglement because neither Alice nor Bob can control their own individual results. Quantum theory is an essentially probabilistic theory, whatever essential might signify here. Even though we had complete knowledge of the state of a quantum system, only conditional probabilities can be predicted in a concrete experiment. All this has been well tested in experiments carried out with particles equally prepared that enter into the apparatus one by one. The individual results occur without sufficient reason: different effects can follow exactly the same causes \cite{PozTom}.\\

As for the other two principles, separability and locality, we need to make a brief digression to try to make them now more precise. To define the separability and locality principles the best words come from Einstein:\\

"If one asks what, irrespective of quantum mechanics, is characteristic of the world of ideas of physics, one is first of all struck by the following: the concepts of physics relate to a real outside world, that is, ideas are established relating to things such as bodies, fields, etc., which claim a "real existence" that is independent of the perceiving subject [\dots] It is further characteristic of these physical objects that they are thoughtless arranged in a space-time continuum. An essential aspect of this arrangement of things in physics is that they may claim, at a certain time, to an existence independent of one another, provided these objects "are situated in different parts of space". Unless one makes this kind of assumption about the independence of the existence of objects which are far apart from one another in space [\dots] physical thinking in the familiar sense would not be possible [\dots]\dots The following idea characterizes the relative independence of objects far apart in space (A and B): external influence on A has no direct influence on B; this is known as the "principle of locality" [\dots] If this axiom were to be [\dots] abolished [\dots] the postulation of laws which can be checked empirically in the accepted sense, would become impossible" \cite{BorEinLet}.\\

The separability principle is violated by entanglement because, as we have illustrated in our example, particles that are far apart from one another may not claim an existence independent from the other. Locality is violated because the results Alice gets here in London have an irretrievable instantaneous consequence there, in Rome.\\

With respect to the more dubious possible violation of the flow of time the basic reference here is the delayed choice proposed for the first time by J. A. Wheeler in 1978 \cite{Whe83a} in an experimental context. Note that these kinds of experiments are not based in the time evolution of the quantum systems (Schr\"{o}dinger equation, for example), but in the measurement postulates. In the original proposal, a quantum particle, a photon let us say, inside a Mach-Zehnder interferometer commits itself to path 1, to path 2 or both before the experimenter chooses the final device (the decision is delayed until the photon has well past the first beam splitter in the interferometer). Broadly speaking, it is as if the photon could know in advance which kind of observation is going to be made. Some authors write it in a more dramatic form: the arrival of the photon will be at detector 1 or at detector 2 of the experimental device depending on if we later learn more information \cite{KimKulShiScu00a}. These delayed choice experiments have been faithfully carried out by different groups recently with convergent results \cite{VarExp3} that once more are in perfect agreement with quantum mechanical predictions, as originally was pointed out by J. Wheeler.\\

The delayed choice experiments have been theoretically analysed also in the context of entanglement \cite{Per99a} and it has been claimed that "even the degree to which quantum systems were entangled can be defined after they have been registered and may even not exist anymore" \cite{BruAspZei05a}.\\

 Henceforth, the conclusion seems to be that the flow of time as it is conceived in Classical Physics is questioned both experimentally, that is, in the delayed choice experiments realized, and theoretically, that is, by entanglement: "in theory, particles can be entangled after their entanglement has already been measured" \cite{Sei05a}. Its meaning should be again reconsidered.\\

This violation of the principles implies an absolute break with the development of human thought. To see the volte-face that quantum entanglement makes in the methodological principles of physics we repeat now frame \ref{fram1} with some necessary adding:

\begin{table}[h!]
\begin{tabular}{|c|c|c|c|c|c|}\hline
 &    Realism & Causality &  Separability &   Locality  &  Flow of time\\\hline
Magic &    yes &    yes &    yes &    yes &  yes\\\hline
Aristotelian physics &     yes  &   yes  &   yes &    yes &  yes\\\hline
Classical physics  &   yes  &   yes  &   yes  &  yes/ (no)&  yes\\\hline
Relativity  &  yes  &   yes &    yes &    yes &  questioned\\\hline
Quantum entanglement &     no &     no &     no  &    no &   questioned\\\hline
\end{tabular}
\caption{Frame 2. - Methodological principles working in magic, Aristotle's physics, classical physics, relativity and quantum physics. The yes/no, etc. inside the boxes are the answers to the question: Does this body of knowledge (magic/ c. physics, etc.) satisfy the principle of \dots?}
\end{table}

\section{Conclusions}
As it can be seen in the previous frame, the break that quantum mechanics introduces in the basic underlying principles that have been working through history in the human thought since immemorial times, is absolute. Our thesis here, presented as conclusion, is that it is precisely the violation of these principles handed down from the remote antiquity and deeply engraved in our minds, what makes quantum mechanical phenomena so surprising that they seem to resemble the results of magic. On the other hand, quantum mechanics is a theory that although it does not satisfy those principles, nevertheless allows us to predict the results of a great variety of phenomena. The break with the principles and the success in its predictions are the support of Greenberger's famous sentence.\\

Entanglement challenges our fundamental concepts about realism, cause and effect, individuation and distance, as well as forcing us to reconsider once more the meaning of time. But it does not challenge at all the Popperian scientific method mentioned above. As we have seen, different theories about nature can be built up using the same \emph{a priori} principles. Can we make the same theory using different \emph{a priori} principles or no principles at all? The role that different \emph{a priori} principles may play in the construction of sciences is a very interesting point, but it will make this paper too long to discuss it in detail. Let us only say that in our opinion the answer to the previous query is in part negative \cite{CommentBohm}. Bell's theorem about the incompatibility between local realist theories, those that would satisfy all previous quoted principles, and quantum mechanics could be a good example to recall in this respect.\\

Quantum mechanics is an extremely well established and confirmed scientific theory, as its capacity to organize a vast field of phenomena demonstrates. But at the same time it is a radically new sort of theory about nature. Based on which we have already said, if it were possible to introduce a quantitative distance measure between different bodies of knowledge, the distance between magic and quantum mechanics would result being the greatest one. Any other scientific theory would result more close to magic than quantum theory. Hence, although the behaviour of quantum systems are "strange" from the point of view of theories that satisfy the above-mentioned fundamental principles, just for this reason of "distance" we should refrain from using words like weirdness or magic in the context of quantum mechanics.\\

There is another interesting conclusion to reach in relation with the first similarity between magic and sciences mentioned above. The idea is the following. Magic and "classical" sciences were based upon the belief of a certain preestablished order in nature. Since the dawn of humankind we have always tried to survive taking advantage of the order, perceived as external and independent of our own activity. In both cases, the origin of the order was some immutable laws conferred by God, Nature or the Big Bang at the beginning of time. God, Nature, the Big Bang, etc., were conceived as external to us and we humans were passive observers that have nothing to do with the order that those everlasting laws continuously produce in nature. Thanks to them we can predict the past and the future. Being they conferred from the beginning and forever, our role as outside observers could only be to discover them. If we discover bad laws, as in magic for example, then we are making a spurious science. But if we discover the good ones, then we are doing genuine science. The scientific method would consist in raising the veil of Maya and gradually uncovering the real essence of the phenomena, the objective reality, the "unalterable givens" (to use Einstein's words, see below). This common image is not only very powerful, but also strong and attractive to the point that many physicists do indeed believe that the equations of physics are obeyed by material reality. However, if we look back to the history of sciences and we analyse any concrete case, it would be clear that this image could not be literally accurate. Think for example in Kepler's laws. We know that they work for short times and in the two-body approximation; they are given in a plane, while the heavenly bodies move in a three-dimensional space (why three-dimensional?); etc. History of physics shows that with the unique exception of current laws and theories, all previous hypotheses have been surpassed by the new order introduced and that, subsequently, they have been proved wrong or limited in some way or another. In fact, none of them last forever \cite{Smo06a}. Are we really only raising the veil of Maya?\\

Quantum Mechanics has definitely changed the role we play in such scenario. By putting us back in the frame as observers, and restoring the essential role that we have always played in the consecution of the order introduced in material reality, quantum mechanics has opened a completely different perspective more adjusted to the facts than the previous one. Let us make a conjecture of how natural laws and theories could be understood from this new perspective. A plausible explanation could be the following. Since the mists of time, humans have operated first with their hands and then with their technological devices with material reality, whatever this would be. Undifferentiated material reality was then decomposed into pieces and composed once again differently from the inside, because we observers are self-organized matter. We began to hold pieces of matter, to "manipulate" them (bones and stones as weapons, for instance), to make material operations with our surrounding material reality. It was this manipulation towards survival what, in the becoming of time and over millions of years, gave rise, from that undifferentiated material reality, to pieces of matter, to something that we now call "material objects". It was then through this manipulation with material objects that some relations, that is, objective recursive relations, were established. What happened later was that these relations could be used to build up new material objects and new relations, and so on. Undifferentiated material reality gave way to pieces of matter, then to material objects and finally to relations between material objects. This everlasting process introduced what could be understood as order within the material reality.\\

It was at a later stage in the development of this process, that we perceived how some relations belonging to a concrete level could be classified and summarized in a certain algorithm able to reproduce all relations of that class. This was called "a natural law". From this new "quantum mechanical perspective", scientific laws appear to be our own constructions: computational algorithms that allow us to condense and reproduce an enormous variety of relations. Magic and classical physics perspectives presented the order as the consequence of immutable laws, while quantum mechanics one presents laws as the consequence of the previous order that we human observers have introduced. There is here a natural "essential tension" between two different aspects. On the one hand, laws are consequences of the order we have managed to introduce in a concrete field of human activity. Yet, on the other, this ordering is independent of our individual will. Matter has its own legality, revealed by the fact that not everything goes. However, this material legality manifests itself through our interventions. It has always been the case, even if for a long time we humans have been able to remove our interventions completely. The phenomena were then explained as if we were "external observers" and as if the mentioned \emph{a priori} principles were "necessary preconditions". But after the quantum revolution we cannot think anymore about the laws as if they had an independent existence of the conceptualization process from which they were built up. Einstein was conscious of the relevance of this forgotten process: "concepts that have proven useful in ordering things easily achieve such authority over us that we forget their \textbf{earthly origins} and accept them as unalterable givens" (emphasis ours) \cite{Ein16a}. And Niels Bohr repeatedly insisted: "from now on the purpose is not to disclose the real essence of the phenomena but only to track down, in so far as it is possible, relations between the manifold aspects of our experience" \cite{Boh34a}. However, it was John Wheeler whom once again was able to make the idea more precise introducing the concept \textbf{participative} \cite{Whe96a}.  What we have tried to show in the previous paragraphs is that the universe we live in \textbf{has always been a participative universe}.\\

As we have outlined, the evolution of the human thought has been from survival to magic, from magic to religion and from religion to science. Nothing in the history of humankind indicates that we are at the end of the road. And, in the same way that four thousand years ago sciences as we conceive of them today were unthinkable, it could result for us unthinkable to conceive the new structure of knowledge that could prevail, let us say, in two or three thousand years from now. In historical terms we are in the path that the following scheme suggests:

  \centerline{$\to$\ Magic\ $\to$\ Religion\ $\to$\ Sciences\ $\to$ ?}

\begin{center}
\epsfig{file=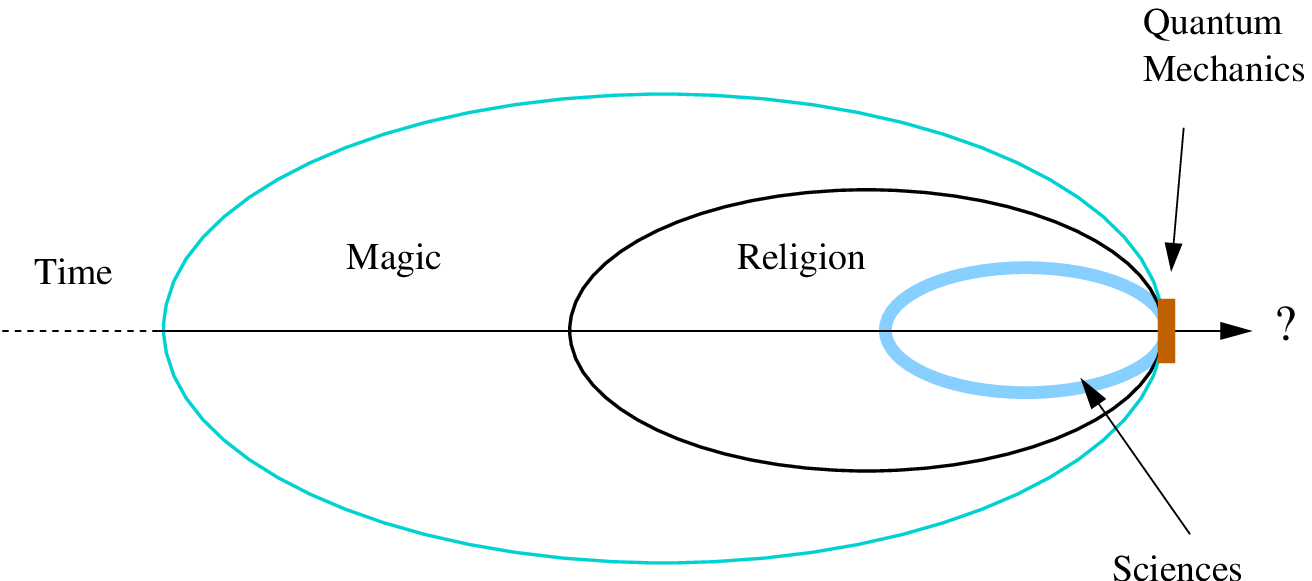,width=12cm}
\end{center}

"The advance of knowledge is an infinite progression towards a goal that for ever recedes" \cite{Fra713}. The interrogation in the figure could be or could be not only an empty symbol. We do not know. But our conjecture is that quantum mechanics constitutes a new form of relation with nature that could finish by giving sense to this question mark.

\section*{Acknowledgments}
The authors would like to thank Juan Le\'{o}n for many stimulating discussions and Emilio Santos, Ram\'{o}n Lapiedra, and Silvia Prost for their comments and the critical reading of the manuscript. This study was supported by the Spanish Ministry of Education and Science under project  No. FIS2008-00288.


\begin{thebibliography}{27}
\expandafter\ifx\csname natexlab\endcsname\relax\def\natexlab#1{#1}\fi
\expandafter\ifx\csname url\endcsname\relax
  \def\url#1{{\tt #1}}\fi

\bibitem{EngDicHisPri88a}
{\em The Shorter Oxford Dictionary on Historical Principles}, Vol. I, p. 1257 (1988).

\bibitem{Kla72a}
J.E. Klauder (ed.)
\newblock {\em Magic without Magic}. W.H. Freeman, 1972.

\bibitem{Tol72a}
J. Toll, page 10 of reference \cite{Kla72a}.

\bibitem{Gre}
D. Greenberger, as quoted by D. Mermin in reference \cite{Mer85a}.

\bibitem{Mer85a}
N.D. Mermin,
\newblock Is the moon there when Nobody looks? Reality and the Quantum Theory.
\newblock {\em Phys. Today} \textbf{38}, 38-47, April (1985).

\bibitem{Gre95a}
D. Greenberger in
\newblock D. Greenberger and A. Zeilinger (eds.),
\newblock {\em Fundamental Questions in Quantum Mechanics: A Conference to Held in Honor of Professor John A. Wheeler}.
\newblock {\em Ann. New York Acad. Sci.} \textbf{755}, p. xiii (1995).

\bibitem{Ard05}
M. Ardnt \emph{et al.},
\newblock Quantum Physics from A to Z.
\newblock {\em arXiv quant-ph/0505187v4.}

\bibitem{VarExp}
See P.G. Merli, G.F. Missiroli and G. Pozzi. \emph{Amer. J. Phys.}  \textbf{44}, 306 (1976) and A. Tonomura, J. Endo, T. Matsuda, T. Kawasaki, H. Ezawa. \emph{Amer. J. Phys.} \textbf{57}, 117 (1989) for experiments carried out with electrons; J. Summhammer, G. Badurek, H. Rauch, U. Kischko, and A. Zeilinger, \emph{Phys. Rev. A} \textbf{27}, 2523 (1983) for an experiment carried out with neutrons; O. Carnal and J. Mlynek, \emph{Phys. Rev. Lett.} \textbf{66}, 2689 (1991) for an experiment carried out with atoms and M. Arndt \emph{et al.}, \emph{Nature} \textbf{401}, 680 (1999) for an experiment carried out with molecules.

\bibitem{Whe83a}
J.A. Wheeler in J.A. Wheeler and W.H. Zurek (eds.) {\em Quantum Theory and Measurement}. Princeton University Press, Princeton, 1983.

\bibitem{VarExp2}
A representative sample of the bulk could be: J. Freedman and J. F. Clauser, \emph{Phy. Rev. Lett.} \textbf{49}, 938 (1972); A. Aspect, P. Grangier and G. Roger, \emph{Phy. Rev. Lett.} \textbf{47}, 460 (1981); P. G. Kwiat , K. Mattle, H. Weinfurter, A. Zeilinger, A. V. Sergienko and Y. H. Shih, \emph{Phy. Rev. Lett.} \textbf{75}, 4337 (1995); G. Weihs, M. Reck, H. Weinfurter and A. Zeilinger, \emph{Phys. Rev. Lett.} \textbf{81}, 5039 (1998); M. A. Rowe, D. Kielpinski, V. Meyer, C. A. Sackett, W. M. Itano, C. Monroe and D. J. Wineland, \emph{Nature} \textbf{409}, 791 (2001); N. Gisin, G. R. W. Tittel and H. Zbinden, \emph{Rev. Mod. Phys.} \textbf{74}, 145 (2002); Y. Hasegawa et al., \emph{Nature} \textbf{425}, 45 (2003); D. N. Matsukevich, T. Chaneliere, S. D. Jenkins, S.-Y. Lan, T.A.B. Kennedy, A. Kuzmich, \emph{Phys. Rev. Lett.} \textbf{96}, 030405 (2006); D. N. Matsukevich, P. Maunz, D. L. Moehring, S. Olmschenk, and C. Monroe, \emph{arXiv:quant-ph/0801.2184} (2008). Note that not all scholars would agree with the interpretation we have made of these experiments. See for example E. Santos, \emph{arXiv:quant-ph/0801.1572} (2008), and references therein.

\bibitem{Fra93a}
James Frazer,
\newblock {\em The Golden Bough. A Study in Magic and Religion}.
\newblock Wordsworth Editins Ltd., London, 1993 (original from 1922).

\bibitem{Frabis}
J. Frazer, in reference \cite{Fra93a}, page 52. Later on Frazer goes as far as saying that "magic directly prepares the way for science. Alchemy leads up to Chemistry", page 92.

\bibitem{Jan04a}
A. Janiak (ed.),
\newblock {Newton: Philosophical Writings}.
\newblock Cambridge University Press, Cambridge, 2004, page 102.

\bibitem{Janbis}
See reference \cite{Jan04a}. Newton speculations about the characteristics that ether might have can be seen in query 21 of \emph{Optics}. As a possible alternative to restore locality, Newton also considered God as intermediary of gravitational interactions. See A. Janiak, \emph{Newton's Philosophy}. Stanford Encyclopedia of Philosophy, 2006.

\bibitem{CommentNewton}
Newton wrote down some experimental notebooks that he kept for about 30 years. One of these documents on alchemy, which Newton likely wrote in the mid-1670s, is an eight-page manuscript now housed at Yale University.

\bibitem{EinPodRos35a}
A. Einstein, B. Podolsky and N. Rosen.
\newblock Can quantum-mechanical description of physical reality be considered complete?
\newblock {\em Phys  Rev.} \textbf{47}, 777 (1935).

\bibitem{Ref11}
Reference \cite{Fra93a}, page 712.

\bibitem{Ref11bis}
Reference \cite{Fra93a}, page 11.

\bibitem{Ref11tris}
Reference \cite{Fra93a}, page 92.

\bibitem{Sch35a}
E. Schr\"{o}dinger,
\newblock The present situation in quantum mechanics.
\newblock {\em Proc. Amer. Phil. Soc.} \textbf{124}, 323 (1935). See also reference \cite{Whe83a}.

\bibitem{Mau07a}
T.W.E. Maudlin,
\newblock Completeness, supervenience and ontology.
\newblock {\em J. Phys. A} \textbf{40}, 3151 (2007).

\bibitem{Mer98a}
N.D. Mermin,
\newblock What is quantum mechanics trying to tell us?
\newblock {\em Am. J. Phys.} \textbf{66}, 753 (1998).

\bibitem{BraMet07a}
G. Brassard and A.A. Methot.
\newblock Can quantum-mechanical description of physical reality be considered incomplete?
\newblock {\em arXiv:quant-ph/0701.001 (2007)}.

\bibitem{Fra07a}
However, there are other tasks that could be carried out faster than the speed of light. See J. D. Franson, Superluminal Generation of Entanglement, {\em arXiv: quant-ph/0704.1468 (2007)}.

\bibitem{CommentKS}
It is possible to maintain some kind of realism by modifying the previous definition by introducing contextual properties. This in essence what the famous Kochen-Specker theorem states. This discussion lies beyond the purposes of this paper.

\bibitem{PozTom}
The clicks registered in a detector that controls a radioactive source are described by a truly random Poisson process. See also, in this respect, the pictures shown by Pozzi \emph{et al.} and Tonomura \emph{et al.} in reference \cite{VarExp}.

\bibitem{BorEinLet}
{\em The Born-Einstein Letters},
 Macmillan, London, 1971, pages 170-171.

\bibitem{KimKulShiScu00a}
Y. Kim, R. Yu, S.P. Kulik, Y.H. Shih, and M. O. Scully. \emph{Phys.  Rev. Lett.} \textbf{84}, 1 (2000).

\bibitem{VarExp3}
For the first realizations of the experiment see: T. Hellmut, H. Walther, A. G. Zajonc, and W. Schleich, \emph{Phys. Rev. A} \textbf{35}, 2532 (1987) and J. Baldzuhn, E. Mohler, and W. Martienssen, \emph{Z. Phys. B} \emph{77}, 347 (1989). The more recent ones are: V. Jacques, E. Wu, F. Grosshans, F. Treussat, P. Grangier, A. Aspect and J.-F. Roch, \emph{Science} \textbf{315}, 966 (2007). See also \emph{arXiv:quant- ph/0703255}; \emph{arXiv:quant- ph/07102597} and \emph{arXiv:quant- ph/0801.0979 (2008)}.

\bibitem{Per99a}
A. Peres.
\newblock Delayed choice for entanglement swapping.
\newblock {\em arXiv:quant-ph/9904042.}

\bibitem{BruAspZei05a}
C. Brukner, M. Aspelmeyer and A. Zeilinger.
\newblock {\em Found. Phys.} \textbf{35}, 1909 (2005).

\bibitem{Sei05a}
See C. Seife, \emph{Science} \textbf{309}, 98, July (2005).

\bibitem{CommentBohm}
Bohm's theory, which satisfies all principles except locality, is in fact a different theory able to reproduce most of the quantum mechanical results. Nevertheless, we should note that some recent experiments testing the de Broglie-Bohm theories against Quantum Mechanics give results that coincide with the standard quantum mechanical predictions and poses a strong constrain on the validity of the de Broglie-Bohm theories. See G. Brida, E. Cagliero, G. Falzetta, M. Genovese, M. Gramegna and C. Novero, \emph{J. Phys. B} \textbf{35}, 4751 (2002) and \emph{Phys. Rev. A} \textbf{68}, 033803 (2003).

\bibitem{Smo06a}
Lee Smolin, Do the laws of nature last forever? \emph{New Scientist}, Issue 2570, pp 30-35, 21 Sep 2006.

\bibitem{Ein16a}
A. Einstein. {\em Phys. Zeitsr.} \textbf{17}, 101 (1916).

\bibitem{Boh34a}
Niels Bohr. \emph{Atomic Theory and the Description of Nature}, page 18. Cambridge University Press, Cambridge, 1934.

\bibitem{Whe96a}
J. A. Wheeler. {\em At Home in the Universe}. Springer-Verlag, New York, 1996.

\bibitem{Fra713}
J. Frazer, in reference \cite{Fra93a}, page 713.

\end{thebibliography}

\end{document}